# Protein Evolution as a Complex System

Barnabas Gall*[1,2], Sacha B. Pulsford*[1,2], Dana Matthews[1,2], Matthew A. Spence[1,2], Joe A. Kaczmarski[3,4], John Z. Chen[1,3], Mahakaran Sandhu[1,2], Eric Stone[5], James Nichols[5], Colin J. Jackson[1,2,3,4]

[1]Research School of Chemistry, Australian National University, Canberra, ACT 2601, Australia

[2]ARC Centre of Excellence for Innovations in Peptide & Protein Science, Research School of Chemistry, Australian National University, Canberra, ACT 2601, Australia

[3]ARC Centre of Excellence for Innovations in Synthetic Biology, Research School of Chemistry, Australian National University, Canberra, ACT 2601, Australia

[4]Research School of Biology, Australian National University, Canberra, ACT 2601, Australia

[5]Biological Data Science Institute, Australian National University, Canberra, ACT 2601, Australia

* These authors contributed equally.
To whom correspondence should be addressed: colin.jackson@anu.edu.au

**Abstract**: Protein evolution underpins life, and understanding its behavior as a system is of great importance. However, our current models of protein evolution are arguably too simplistic to allow quantitative interpretation and prediction of evolutionary trajectories. Viewing protein evolution as a complex system has the potential to advance our understanding and ability to model protein evolution. In this perspective, we discuss aspects of protein evolution that are typical of complex systems, from nonlinear dynamics, sensitivity to initial conditions, self-organization, and the emergence of order from chaos and disorder. We discuss how the growth in sequence and structural data, insights from laboratory evolution and new machine learning tools can advance the study of protein evolution and that by treating protein evolution as a complex adaptive system, we may gain new insights into the fundamental principles driving biological innovation and adaptation and apply this to protein engineering and design.

# GLOSSARY

**Analytical solution**: An equation directly returning the variables at any time t, without further computation. For example, the sequence after ten generations, given the initial sequence.

**Attractor**: A state in a dynamical system that the system evolves towards over time. Attractors can take different forms, such as fixed points, limit cycles or strange attractors.

**Bifurcation**: A change in a system's dynamic structure as a control parameter is varied, leading the system into new behaviors ('pathway to chaos').

**Chaos theory**: The study of deterministic dynamic systems which are unpredictable due to high sensitivity to initial conditions.

**Complex system**: Systems composed of many interacting components (and their environment) that can exhibit elements of chaotic behavior but do not necessarily conform to a strict deterministic definition of chaos.

**Deterministic process**: A process in which a given input will consistently return the same outcome.

**Disorder**: Inherent unpredictability or randomness in the structure or behavior of a system.

**Non-linear dynamics**: When elements in a system interact with each other non-additively to determine the behavior of the system.

**Periodicity**: Repeating patterns of cycles that arise from complex dynamics.

**Phase space**: A representation of all possible states in a system and all possible forces acting upon the state (e.g. mutation and selection).

**Self-organization**: When order within a system arises from interactions within the system alone without external factors.

**Strange attractor**: An attractor found within a system that has a fractal structure, e.g. due to bifurcation through evolution.

**Chaos and complex systems**. Chaos theory and the study of complex systems represent relatively modern scientific frameworks, both emerging as prominent fields in the latter half of the 20th century[1–6]. However, there has been interest in the idea of order from chaos since at least the 8th century BCE — as Hesiod wrote in his Theogony, "...first Chaos came to be"[7]. Chaos theory addresses deterministic systems in which infinitesimal variations in initial conditions lead to divergence in trajectories over time. This unpredictability is exemplified by systems such as the double pendulum, where seemingly minor initial differences can lead to vastly different outcomes as time progresses[8]. Chaos theory thus provides a lens for understanding how seemingly random behaviors can emerge in systems that are fundamentally deterministic. Complex systems are broadly defined as systems with many interacting elements. Critically, the collective behavior of a complex system emerges from intricate networks of nonlinear interactions and cannot be understood by examining their individual components in isolation[6,9]. A complex system may exhibit chaotic tendencies, but chaos theory does not encapsulate all complex systems nor the behaviors therein. Examples of complex systems include biological ecosystems[10], economies[11,12], neural networks[13], and social systems[9] — each comprising a multitude of interdependent components that interact to produce unpredictable, emergent behaviors.

Complex systems are distinguished by several core characteristics[6,9,14]. They are inherently nonlinear, meaning that their response to inputs is not proportional, and small changes can produce disproportionately large effects. This nonlinearity contributes to the system's potential for sudden shifts or transitions in behavior. Another hallmark is self-organization, where ordered structures or behaviors spontaneously emerge from local interactions without centralized control. Complex systems often exhibit fractal patterns, where self-similar structures appear across different scales, indicating underlying organizational principles that recur within the system's hierarchy[15,16]. Feedback loops are also prominent, which introduce recursive processes where the output of one part of the system influences other parts and can either stabilize or destabilize the system, allowing it to adapt to internal and external changes[16]. Finally, entropy, a concept often associated with disorder, also plays a role in understanding the evolution of complex systems; while entropy typically increases in complex systems over time, localized decreases in entropy can occur through processes of self-organization[16,17]. This interplay between entropy and self-organization is fundamental to the dynamic evolution of complex systems and contributes to their resilience and adaptability. These characteristics of complex systems provide a valuable framework for investigating protein evolution, a domain inherently complex in its molecular interactions, adaptive processes and changing selection pressures.

Many of the techniques and concepts at the heart of complex systems theory have been applied to evolutionary biology. For example, Fisher's Fundamental Theorem draws parallels between evolution and the second law of thermodynamics, and, paired with his 'Fisher information' to measure indeterminacy, established a foundation for studying entropy and disorder in evolutionary processes[18–

[20]. Similarly, the diffusion approximation, independently developed by Wright and Kimura, bridges both statistical physics and population genetics to model the effect of drift, selection and mutation on allele frequencies[21–23]. Mathematical models of evolution, primarily in the field of population genetics, have increasingly drawn on these ideas and other concepts as the fields of nonequilibrium thermodynamics and statistical mechanics developed[24–26]. Additionally, statistical mechanics techniques such as path ensembles, often used to study complex systems, have been successfully applied to examine fitness landscape topologies[27–29]. While most of these examples are from population genetics, similar principles can also be applied to molecular evolution. Viewing protein evolution as a complex system allows us to draw upon and expand these concepts, refining our understanding of the forces that govern protein evolutionary trajectories. This is particularly timely given the rise of *in silico* evolution and other computational approaches to modeling protein evolution.

**Protein evolution is a complex system.** In protein evolution, mutation and selection act as forces that change amino acid sequences over time[30]. Identifying and understanding the interactions of these components is central to describing how this system evolves. In a purely deterministic evolutionary system, one could theoretically predict the precise trajectory of a protein's evolution with perfect accuracy if the current sequence state and selection pressure are known with infinite precision[31,32]. However, as our current understanding of these driving forces and their seemingly intricate interactions are limited, we must model evolution probabilistically, i.e., mutations arise by chance and selection pressures can fluctuate unpredictably. This is compatible with complex systems theory, which accounts for a vast array of interacting variables within the system, many of which we cannot hope to include in modeling.

The specific nature of the evolutionary mechanism in protein evolution is worth defining; it is discrete, with the sequence states changing via (generally) single mutations in a stepwise manner. It is indeterministic in the sense that many alternative future states are theoretically accessible for any given position along a protein sequence (each with their own subset of potential future states), but only one may be actualized at a specific bifurcation time point (Fig 1a). This gives rise to the concept of evolutionary trajectories and the multiplicity of possible 'routes' from any given starting point. Simple models of protein evolution can help illustrate the interplay of selection and mutation forces critical in defining which trajectories are actualized. For instance, consider the starting sequence 'AAAA' (Fig 1b-d). Under low pressure, the system can drift, sampling every state and the trajectories therein. As the stringency of the selection pressure increases, the system becomes constrained and can become "stuck" at a local maximum (e.g 'ABBA' in Fig 1d). Selection pressures can be viewed as a force that makes the otherwise stochastic process of random mutagenesis and drift essentially deterministic at its most stringent levels.

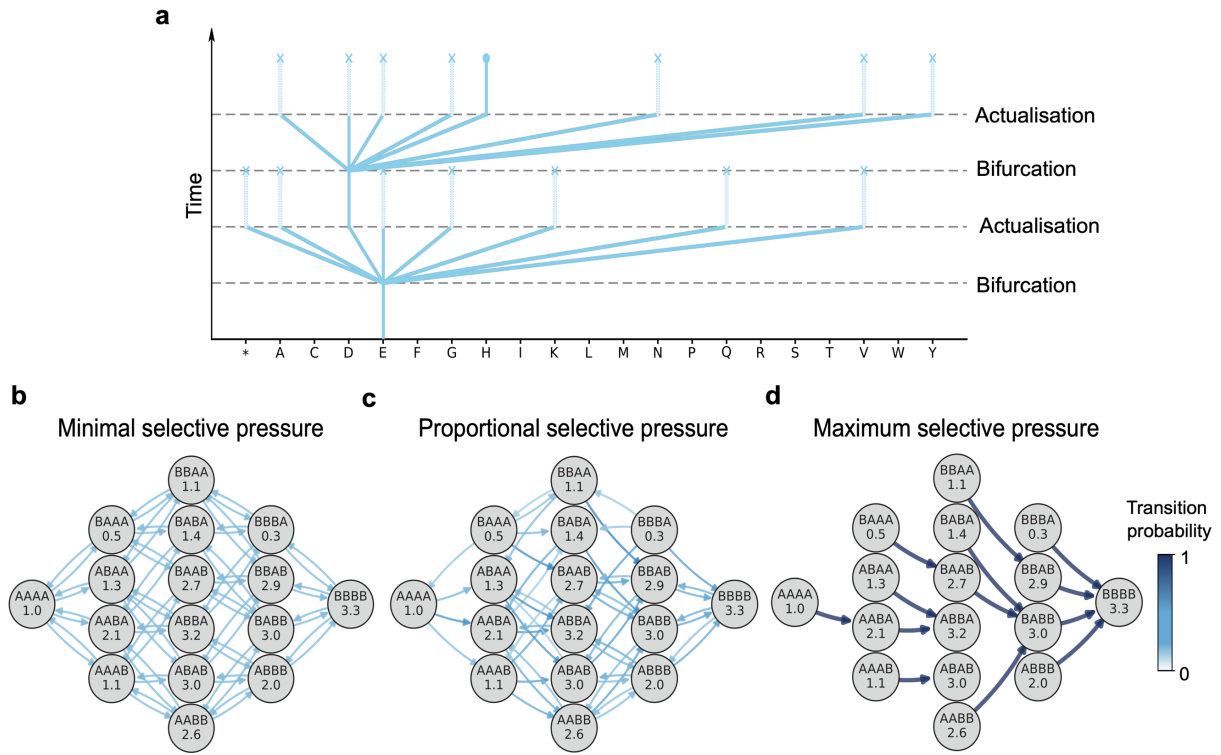

**Figure 1.** (a) Depiction of bifurcation and actualization at a single sequence position, with only nucleotide substitutions allowed. Starting with glutamic acid (GAA), 12 codons (7 amino acids and stop codon(*)) are accessible at the bifurcation point via single base changes. Only aspartic acid is actualized, followed by histidine after the next bifurcation. (b-d) Possible trajectories given a 2-character sequence space spanning 4 positions and associated fitness values under low, proportional and maximum selective pressure. (b) Minimal selective pressure allows all next sequence states. (c) Proportional selective pressure makes increasing sequence fitness more probable. (d) At maximum selective pressure, only sequences that increase fitness are accessible.

In "wild" populations, the underlying selection and mutation forces manifest in more diverse, complicated ways, feeding back on each other to exponentially increase the complexity of the system. For instance, there are typically many compounding selection pressures at play, such as protein stability, catalytic activity, regulatory functions, and expression levels, all of which may interact to create a complex, rugged, landscape[33–35]. Moreover, fitness is often non-stationary, with the system continually cycling through variants in response to changing environmental pressures, interactions between components, or spontaneous mutations within the population itself[36–38]. These oscillations prevent the system reaching a global equilibrium, leading to continuous flux of variants that can make simple predictions challenging. Recombination, frameshifts, indels and other more extreme mutation events can also drastically reconfigure available sequence space[39–41]. In comparison, laboratory directed evolution studies typically impose a smaller number of strong selection pressures under highly controlled conditions, favouring the accumulation of adaptive mutations. This leads to more deterministic outcomes by minimizing the stochasticity of both environmental conditions and mutation rates. Nevertheless, even in laboratory studies, neutral drift and/or multiple selection pressures such as protein stability and regulatory functions may interact to erode predictability of the system[42–44]. Viewing protein evolution as a complex system of a discrete, indeterminate nature could help account for these complicated, interacting factors and enrich our understanding of the system as a whole.

**Initial conditions, contingency and directionality in evolution.** A hallmark of complex systems, and one that is particularly relevant to evolutionary processes, is sensitivity to initial conditions[45]. Over evolutionary timescales, two nearly identical protein sequences can diverge into structures with entirely different functions due to the amplification of small initial differences through the complex interplay of mutation, selection, and chance events. Edward Lorenz famously described this sensitivity to initial conditions in his work on weather patterns, noting that two states differing by small amounts can evolve into considerably different states, making long-term prediction essentially impossible[2]. This concept is often referred to in the protein evolution field as evolutionary contingency and underscores how seemingly insignificant "neutral" differences at the outset of an evolutionary trajectory can lead to dramatically divergent outcomes over time[46–49]. The extreme consequences of bifurcation events is epitomised by deep evolutionary relationships identified between protein families such as the TIM barrel and flavodoxin-like folds that today have seemingly irreconcilable functions and forms[50].

A growing body of work illustrates the context dependency of mutations at the molecular level, wherein the effect of a mutation depends on the genetic background in which it occurs. Epistasis has emerged as a particularly important type of context dependence that is invoked when the combined effect of two or more mutations deviates from that predicted by their additive individual effects. Thus, a mutation may be beneficial or lead to novel activity in one sequence background but have neutral or deleterious effects in another. For example, deep mutational scanning of ancient steroid hormone receptors and close analyses of the laboratory directed evolution of the enzyme phosphotriesterase both illustrate how the effect of chance molecular events radiate, shaping the outcomes of evolution[46,48].

An extension of dependence on initial conditions is irreversibility. Dollo's Law emphasizes that traces of the intermediary states will persist and influence future possibilities[51,52]. This makes the reemergence of identical states highly improbable[51–53]. From a purely statistical perspective, the likelihood of a mutation reoccurring at a specific site is already slim when considered over an entire biological sequence within a finite population size. Epistatic interactions compound this effect. For instance, a neutral (and thereby reversible) mutation may become entrenched by a subsequent restrictive substitution that renders the ancestral state deleterious. This was illustrated in the deep sequencing of substitutions accumulated in the long-term evolution of the eukaryotic heat shock protein Hsp90[54]. Here, many reversions to ancestral states were deleterious, revealing a daisy-chain effect wherein a permissive mutation becomes entrenched and irreversible as a mutation contingent upon it occurs that, in turn, permits a subsequent substitution, and so on. Each change closes reverse paths at some sites and opens forward paths at others. Extreme examples of this ratchet-like action lead to the fixation and irreversibility of activities limiting access to 'adaptive peaks'. This was observed during the evolution of hormone receptor specificity, and in studies that reversed the laboratory directed evolution of enzymes[55,56]. While reverse evolution is not theoretically impossible, entrenchment and irreversibility appear to be the pervasive force in protein evolution. In these ways, the effects of chance acting on minute differences results in extreme effects on evolutionary trajectories, shaping life's diversity and propelling evolution forwards.

**Turbulence, entropy and self-organization.** The chaotic trajectories of protein evolution, especially under low selective pressure where neutral drift is prevalent, bear striking resemblance to turbulence in complex systems theory — high entropy states of seemingly random, disordered behavior where small changes can propagate into large, unpredictable outcomes[57]. In protein evolution, this turbulence could manifest in a high variability of sequences across time, with diverse variants coexisting within a population at comparable frequencies. Even when an evolutionary system appears to be in equilibrium, shifts in selection pressures can disrupt this balance, triggering turbulent dynamics that drive the system toward a new state. An extreme example could be the de novo emergence of proteins, where translated protein sequences under no selection pressure can drift and sample a multitude of states before function emerges[58,59].

As in turbulent flows, the system may eventually settle into a new low entropy state, dominated by the fittest sequences. However to sustain such a change, a continuous input of force — in this case, selective pressure — is required, with diminishing returns gradually stabilizing the system[60]. Despite its apparent randomness, turbulence often reveals emergent patterns and structures, uncovering a hidden order within disorder[14]. In protein evolution, this is exemplified by the emergence of novel folds and functions — metaphorical "islands of stability" amidst turbulent sequence drift[44,59]. These patterns are increasingly observed not only in natural evolution but also in protein design[61–63]. Prigogine's foundational work on dissipative systems and self-organization offers a compelling framework for understanding these dynamics[64]. When systems are pushed far from equilibrium, they can exhibit surprising phenomena, giving rise to new forms of order.

In the same way that a fluid stream can transition from laminar to turbulent flow, protein structures can exist in ordered or disordered states as determined by their sequence[65–67]. For example, evolutionary adaptation is underpinned in many instances at the biophysical level by conformational sampling across the protein conformational landscape, whereby new activities may emerge from a promiscuous, and conformationally flexible, intermediate[68]. One example of this is the evolutionary transition between enzymes specific for arylesters or phosphotriesters, which proceeds through disordered intermediate states that can sample conformations optimized for both substrates[56,60]. As selection pressure is exerted in either direction, diminishing returns and stabilization of the sequence and structure ensues, as is typical for turbulence in complex systems theory. This is extended by examples of fold switching proteins. In this case, if sequences encoding different folds are sufficiently close in sequence space, a protein may pass through a brief period of turbulence and disorder to adopt a new fold[69]. Recent work has shown fold-switching is far more prevalent than previously suspected and that sequence features that predispose to this can be observed and modeled through deep learning ML approaches[69,70].

A final example of the interplay between order and disorder is seen in the principle of consensus design, exemplified by the well-documented "consensus effect." Here, thermostability can be enhanced in proteins by introducing consensus mutations — amino acid substitutions that align a protein's sequence closer to the consensus or equilibrium sequence derived from a family of homologous proteins[63,71]. Most mutations are inherently destabilizing, with selection acting to eliminate highly unstable proteins; however, strong empirical evidence demonstrates the marginal stability of natural proteins[72–74]. That is, native structures are typically only as stable as their functional environment requires, often existing at the cusp of conformational instability[75,76]. This results in a system where sequences are constantly pulled toward disorder by the dissipative effect of mutational turbulence while selective pressure for functional stability exerts a concurrent counterbalancing force, guiding proteins toward the low-entropy state of a stable fold. Consensus mutations harness this principle, leveraging evolutionary signals to impose order on an otherwise turbulent landscape and thereby stabilizing proteins through the alignment of their sequences with an optimal equilibrium state[71]. This illustrates how the tension between disorder and selective forces can give rise to emergent stability, a hallmark of complex systems.

**Phase space and evolutionary trajectories.** Phase space is a useful concept and tool in the study of complex dynamical systems and can be applied to conceptualize sequence space and evolutionary trajectories therein. It describes a multidimensional space where each dimension is a variable of the system and each point is a unique state of the system[77,78]. We envisage the phase space of protein evolution to be all possible combinations of selection pressures, and their change over time, paired with real sequence space, akin to how the phase space of a pendulum pairs velocities to each angle of the pendulum in real space. In phase space, evolutionary trajectories would be constrained to sequence space pertaining to fold and function, nonfunctional sequences would be stationary points, unable to be entered or exited. If we were to be able to visualize phase space, we would be able to see all possible trajectories of evolution given the initial conditions, much as phase space can be used in visualizing possible trajectories of classic dynamical systems such as the forced pendulum.

The irreversibility of evolution and the accumulation of mutations at each generation significantly affect how variants traverse this space. Applying this to protein evolution builds on the classic fitness landscapes to help visualize how concepts such as turbulence play out in molecular evolution. The underlying "churn" of neutral drift constantly shifts the starting point for potential evolutionary trajectories, while changing environmental selection pressures alter the selection landscape. The result is a pattern of diversification and bifurcation that shapes the observed fitness topologies.

**Strange attractors and fractal geometry.** In complex systems theory, an attractor is a subset of states in the phase space of a dynamical system that the trajectories of the system tend to evolve towards regardless of initial conditions (Fig 2a)[16]. Within the attractor trajectories can be periodic or chaotic, and highly sensitive to perturbation, but are constrained to within the attractor. This provides a powerful framework for understanding self-organization in evolutionary trajectories[79,80]. In the context of protein evolution, native protein folds or fitness maxima (which are not mutually exclusive) can represent types of attractors that sequences in turbulent regions are forced towards through selection pressure (Fig 2b). In theory a fitness attractor could occupy a single, stationary point in phase space that states converge upon, however because many sequences are likely to be similarly active, and it is essentially impossible to select between them at very close fitness levels, in practice a fitness attractor would exist as a distribution of sequences. Fold attractors are non-stationary attractors occupying a broad region of sequence space, exemplified by protein folds being conserved with <30% sequence identity[81].

A strange attractor often exhibits fractal-like dimensionality, i.e recursive processes or feedback loops where simple rules apply repeatedly, leading to intricate patterns that are similar at every scale[16,82]. The most obvious example is the bifurcating structure of phylogenetic trees, where branches continually split into smaller branches, mirroring the fractal patterns observed in physical trees (Fig 2c)[83,84]. This bifurcation is a fundamental aspect of protein evolutionary dynamics. As mutations arise, the driving force underlying the system, they generate bifurcations in a sequence's trajectory, resulting in a pool of variants all representing different initial conditions/states. Importantly, bifurcation in sequence trajectories may also arise from the bifurcation of parameters acting upon the trajectories, such as two populations segregating, and being acted upon through different selection parameters. As these states progress and interact with selection pressures, they trigger irreversible actualization of unique paths through sequence space[32]. The continuous bifurcation of variants at each generation leads to both fractal-like convergence of novel sequences around the functional sequence space area/strange attractor.

The strange attractor concept explains how evolutionary trajectories can appear to converge towards certain structures/function while never exactly repeating due to the sensitivity to initial conditions, directionality and bifurcation, and the functionally infinite size of combinatorial sequence space (Fig 2d). This creates a fascinating dynamic where sequences are simultaneously attracted to certain states (folds or functions) while continually diverging and exploring new space in characteristic fractal-like geometries (i.e. phylogenetic trees)[84]. The infinite-dimensional nature of strange attractors in phase space, contrasted with their finite dimensional manifestation (perhaps analogous to the possibly finite number of stable protein folds), mirrors the duality of theoretical possibilities and constrained realities of evolutionary exploration. Viewing these landscapes through the perspective of dynamic systems helps us understand the aperiodic nature of evolutionary trajectories, where patterns may resemble each other but never exactly repeat, and the complex dynamics that underpin these emergent behaviors.

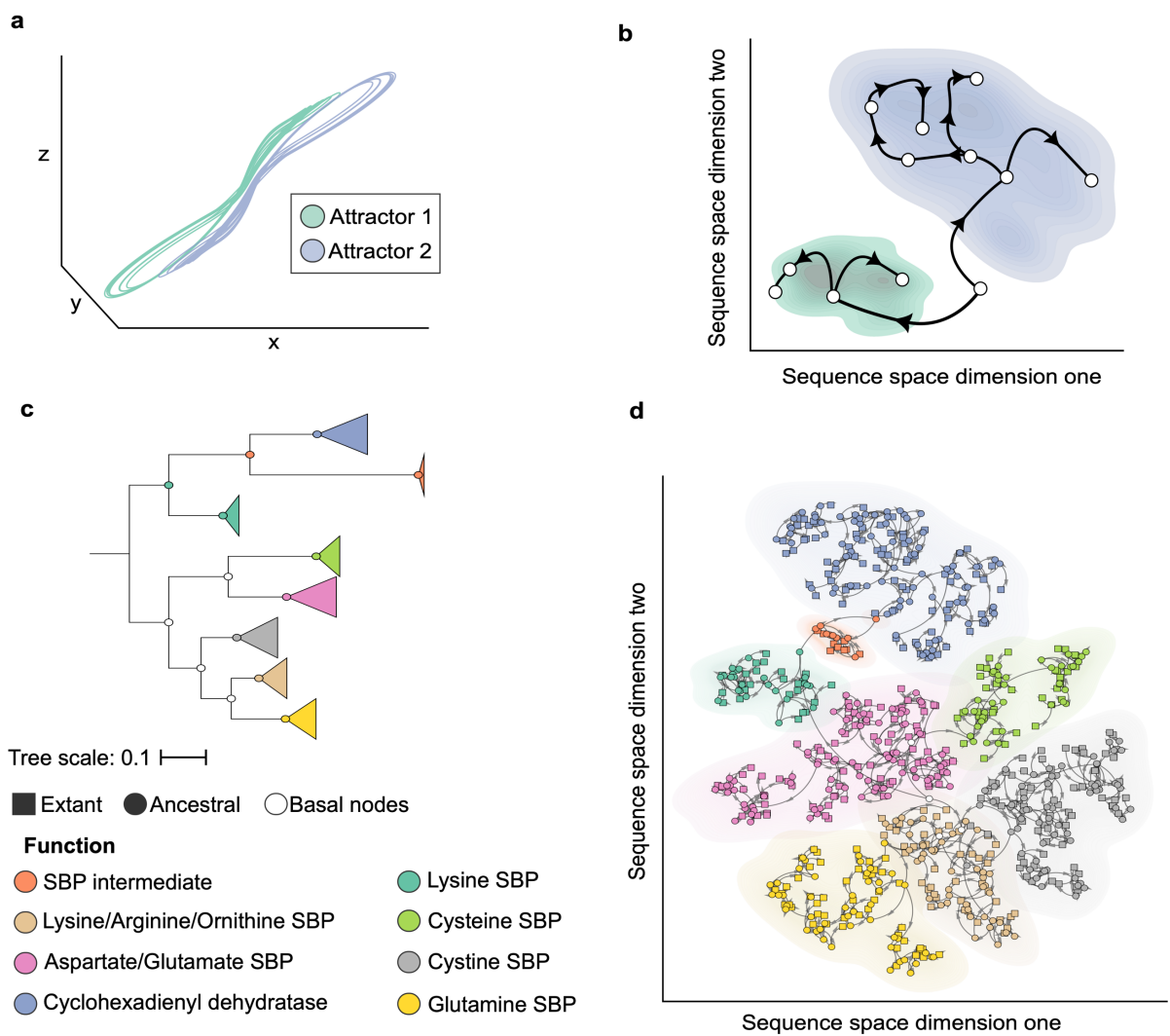

**Figure 2:** (a) Example trajectories Lorenz system showing trajectory bifurcation converging on different strange attractors due to different parameters σ:10, β:8/3, ρ:100 (green) ρ:300 (blue) Initial state (1.0,1.0,1.0) indicated by white point. (b) Bifurcating evolutionary trajectories orbit "strange attractors" pertaining to function in sequence space indefinitely. (c) Phylogenetic tree of solute binding proteins and cyclohexadienyl dehydratase[85]. Clades are colored by function, inferred by characterizing the last common ancestor and known extant sequences. (d) Uniform manifold approximation and projection (UMAP) of one-hot encoded ancestrally reconstructed and extant sequences from the phylogenetic tree

in (c), depicting the evolutionary trajectories in sequence space. Arrows indicate direction of evolution. Functional basins and strange attractors are observed, with isofunctional proteins clustering in similar regions.

**Machine learning in the study of complex systems.** Complex systems pose significant challenges for traditional analytical approaches, even if all variables are known. For example, mapping phase spaces of real-world systems is often computationally intractable due to the sheer scale and intricacy of these interactions. Machine learning (ML) has emerged as a useful tool in this regard, allowing for the behaviors of complex systems to be modeled, analyzed, and predicted, without requiring explicit equations for their dynamics[86,87].

Several approaches have been successfully applied to model complex systems. Among these, deep learning models like graph neural networks (GNNs) and their variants stand out for their ability to identify latent structures and high-dimensional patterns in data where interactions evolve over time[88,89]. One example is leveraging GNNs to simulate and predict the micro-dynamics of systems, while maintaining physical interpretability of interactions[90]. Additionally, physics-informed neural networks (PINNs), which embed known physical laws into the learning process, can enforce constraints to ensure that models remain physically consistent while approximating complex dynamics[91,92]. These have proven effective in modeling both chaotic systems, like the double pendulum, and systems with partial or noisy observations[93,94]. Another approach applied to complex systems with notable success is reinforcement learning (RL), which trains a model through trial and error rather than minimizing residuals[95]. By optimizing decision-making through trial and error, RL can be used to control complex systems, stabilizing chaotic systems and optimizing performance in adaptive environments[86,96,97]. Finally, for dimensionality reduction and phase space exploration, manifold learning techniques such as autoencoders and variational autoencoders (VAEs) help to map high-dimensional system dynamics into low-dimensional latent spaces[98]. These representations aid visualization of phase space structure, enabling identification of attractors and transitions between regimes. Coupled with generative models like diffusion networks, ML can simulate potential trajectories, revealing hidden pathways and behaviors in phase space that were previously inaccessible[99]. Thus, by combining predictive power with interpretability, ML approaches can enhance our understanding of complex dynamics and enable practical applications in forecasting, optimization, and control of complex systems.

**Machine learning methods to model protein evolution.** The application of ML to the study of protein evolution has greatly advanced our ability to model and predict evolutionary dynamics. While many approaches have been deployed, a few key milestone developments exemplify this progress. Most notably, Protein Language Models (PLMs), such as ESM-2, utilize masked language modeling to capture statistical patterns encoded in protein sequences, enabling them to predict functional effects of mutations and generate novel, functional proteins[100–102]. Another interesting application of PLMs is seen in the concept of "evolutionary velocity" which uses PLMs to construct a vector field of possible evolutionary trajectories, mapping the trajectory of proteins through sequence space and offering insights into mutational strategies across diverse timescales[103]. PLMs have been expanded into generative and multimodal/multiscale models like ESM-3, which integrates sequence, structure, and function into a unified latent space, or Evo, which integrates DNA, RNA, and protein data to simulate multiscale evolutionary phenomena to include co-evolutionary and systems-level dynamics[102,104]. A further extension of this technique has been the incorporation of Ancestral Sequence Reconstruction (ASR), a well-established statistical approach for inferring ancestral protein sequences, to generate synthetic datasets that incorporate phylogenetic and evolutionary constraints[105]. Utilizing ancestral sequences appears to enable models to learn smoother fitness landscapes, improving the predictive accuracy of downstream tasks such as fitness estimation and protein engineering as exemplified by local ancestral sequence embedding (LASE)[106].

**Limitations of ML approaches in modeling protein evolution.** Despite significant advancements, modeling protein evolution remains limited by the complexity of its dynamics and the challenges of accurate prediction. Key hurdles include rugged fitness landscapes, nonlinear epistatic interactions, and the stochastic nature of mutations and selection. ML models, such as PLMs, struggle to generalize or extrapolate beyond training data, particularly under variable selective pressures or when predicting novel functionalities[107,108]. Unlike some complex systems like weather or robotics, where governing laws (e.g., Navier-Stokes equations) or well-defined parameter spaces constrain predictions, protein evolution lacks universal "laws" and suffers from sparse, discrete data[109]. Fitness landscapes are difficult to navigate, and the integration of temporal dynamics — spanning geological timescales and diverse environmental pressures — remains a significant obstacle. While promising approaches like bandit theory and evolutionary velocity modeling have emerged, their utility is constrained by limited data quality and scope[110–112]. Overcoming these challenges will require integrating multimodal data, biophysical constraints, and advanced generative techniques to better capture the stochastic and multi-dimensional nature of protein evolution, a system far more complex than many where ML has found predictive success.

Indeed, the ability of PLMs to learn evolutionary patterns has perhaps been overstated in some cases[113,114]. By compressing vast quantities of sequence data, PLMs essentially store coevolutionary information. In this way, PLMs can be seen as stochastic parrots, in that they do not 'understand' the fundamental biophysical nature of proteins or how they evolve to acquire novel functions[101,115–117]. As such, approaches that increase the availability of coevolutionary information provided to the model have been shown to improve model performance in fitness prediction tasks. Indeed, the observation that simple one-hot embeddings can outperform, or be on par with, the encodings of large language models, raises into question the "understanding" of these models[106,118,119]. For example, the green fluorescent protein esmGFP, generated by ESM3 'simulating 500 million years of evolution', was produced using the generative capacity of ESM3 to predict probable sequences for 229 residues to complete the protein after inputting the key backbone coordinates and residues for chromophore maturation, as illustrated in Figure 3a[102]. In other words the model is constructing a likely sequence based on the conditional probabilities, rather than modeling an evolutionary trajectory from the last universal common ancestor. A true approximate analytical model of evolution would be able to extrapolate sequence trajectories into the future.

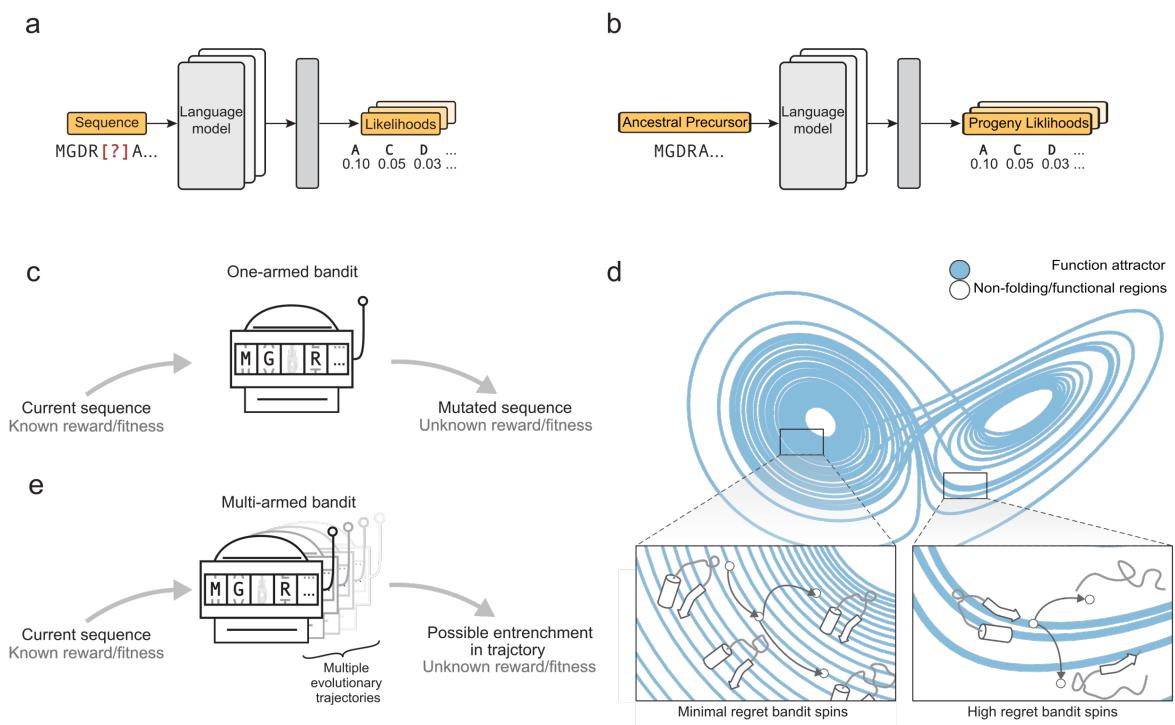

**Fig 3:** (a) Diagram of a protein language model trained using the masked language modeling objective. The model is trained to predict the likelihood of the red 'hidden' token represented by the question mark (b) An example of a language model that may be used for modeling evolutionary trajectories, in this instance the model need predict the next sequence in the evolutionary trajectory (c) A one-armed bandit where a current sequence with a known fitness is mutated, increasing exploration, into a new protein sequence with unknown fitness. (d) The impact of one-armed bandit algorithms on a strange attractor. Maintaining evolution on the attractor (in blue) minimal regret is achieved per spin (mutation) whereas regions of sequence space where there are more non-functional proteins (white space), high regret is risked per spin (mutation). (e) A multi-armed bandit where rather than single mutations, the outcome is entrenchment in new trajectories with unknown pathways and outcomes.

**Future directions: harnessing complex systems theory to advance protein evolution.** Protein evolution is undeniably an extremely challenging system to accurately model. However, by applying underutilized tools from complex systems theory, which are becoming more accessible with increasing computational power and more sophisticated algorithms, rapid advancements may be imminent. For example, bandit theory offers a framework to balance exploration and exploitation in evolutionary modeling, simulating trajectories under varying selective pressures to optimize fitness improvements, and has already begun to be applied in the protein domain[110–112]. Distributed information bottlenecks may be able to quantify how mutations influence global fitness, aiding in the prediction of epistasis and fitness peaks[120]. Aspects of chaos theory can provide insights into evolutionary dynamics by performing post-hoc evolutionary model analysis to identify fitness landscapes as strange attractors, identifying stable regions and transitions between peaks, as have applied to other dynamic systems[121]. Game theory can capture co-evolutionary dynamics, such as host-pathogen interactions, introducing strategic modeling to predict adaptive responses[122]. Agent-based models enable the simulation of collective protein behaviors, such as metabolic pathways, within synthetic ecosystems[123]. High-dimensional optimization techniques, like covariance matrix adaptation, could enhance *in silico* evolution by efficiently navigating sequence space while maintaining constraints. Network theory offers tools to map evolutionary pathways and identify key mutational nodes[124].

Finally, the ability for *in silico* evolution models to simulate the evolution of vast populations, generating millions or even billions of variants, far surpassing the populations typically explored in laboratory-directed evolution, offers an exciting new tool for investigating evolutionary dynamics[125,126]. This scale reduces the stochastic effects of mutation by comprehensively sampling sequence space, diminishing the sensitivity of evolutionary trajectories to random mutations accrued in early generations and making the process more deterministic. Together, these methods promise to make *in silico* evolution the dominant paradigm, enabling exploration of sequence spaces far beyond what is feasible in nature or the lab and driving new frontiers in protein engineering and evolutionary discovery.

**Summary.** In this perspective we have described how protein evolution embodies the defining characteristics of complex systems, including nonlinear dynamics, sensitivity to initial conditions, self-organization, and the emergence of order from chaos. The vast sequence space encoded by genetic material, shaped by the interplay of diverse selection pressures, creates a landscape that is both deterministic in its fundamental principles and unpredictable in its specific evolutionary outcomes. Epistasis further complicates this landscape, with the effects of mutations intricately dependent on their genetic context, forming a web of interactions that drive evolutionary trajectories. The fractal-like branching of evolutionary trees and the presence of strange attractors, such as stable folds or functional states, highlight the deep alignment between protein evolution and complex systems theory. ML offers opportunities to advance our understanding of protein evolution as a dynamic system. By uncovering patterns in large datasets and modeling nonlinear interactions, ML is uniquely suited to tackle the complexity of protein evolution. Emerging tools rooted in complex systems theory, such as bandit theory and phase space analysis, hold particular promise for exploring sequence space and predicting evolutionary pathways. As computational methods grow more sophisticated, they may not only enhance our understanding of evolution but also enable precise modeling and prediction of evolutionary trajectories. This convergence of ML and complex systems theory is poised to further accelerate the ongoing revolution in protein engineering and design.